\documentstyle[11pt,iau185,twoside,epsf]{article}

\begin{document}
\title{Insights from Pulsating Nova Envelopes}
\author{K. Schenker\altaffilmark{1}}
\affil{Theoretical Astrophysics Group,
University of Leicester, Leicester, \mbox{LE1 7RH}, U.K.\\
Astronomisches Institut der Universit{\"a}t Basel, Venusstrasse 7,
CH--4102 Binningen, Switzerland}
\altaffiltext{1}{e-mail: kjs@star.le.ac.uk}

%%%%%%%%%%%%%%%%%%%%%%%%%%%%%%%%%%%%%%%%%%%%%%%%%%%%%%%%%%%%%%%%%%%%%%%%

\begin{abstract}
Based on a linear and non-linear study of radial pulsations in the
envelopes of classical novae (Schenker 1999), I discuss the
% Schenker 1998; 
results both from the point of view of pulsation theory as
well as their consequences for current nova models.

Starting from initially static envelope structures at various stages
during the decline of a nova outburst, strong `running wave'
instabilities have been found that rapidly grow into shocks.
Improved analytical concepts give a new direction to the
interpretation of such highly non-adiabatic radial pulsations.
%A particular point of interest is the relation of these strong
%instabilities to recent discussion of Super-Eddington atmospheres and
%winds (Shaviv 2001) as well as to `strange modes' (Glatzel 1994; Saio,
%Baker, \& Gautschy 1998). 

For direct observational confirmation a search for short period
variability in the UV and soft X-ray is suggested during the very late
decline phase. 
Speculative consequences for mass loss scenarios in novae due to these
instabilities will need some more work in the future. 
\end{abstract}

%%%%%%%%%%%%%%%%%%%%%%%%%%%%%%%%%%%%%%%%%%%%%%%%%%%%%%%%%%%%%%%%%%%%%%%%

\keywords{Stars: oscillations, Stars: novae, cataclysmic variables}

%%%%%%%%%%%%%%%%%%%%%%%%%%%%%%%%%%%%%%%%%%%%%%%%%%%%%%%%%%%%%%%%%%%%%%%%

\section{Introduction}

Classical nova outbursts take place in mass transferring binaries,
when hydrogen-rich material is accreted onto a white dwarf (WD). 
Once enough matter has accumulated, nuclear burning sets in explosively
causing a complex and dynamical initial phase of the nova outburst.
Around visual maximum however, a much more steady phase begins,
where an extremely large ratio of luminosity to mass occurs and the
structure of the star resembles that of a giant: a nuclear burning
shell sitting on top of the WD, covered by an extended envelope of
very little mass (typically $< 10^{-4} \, M_{\sun}$).

Schenker (1999) analyses the pulsational stability of such a
structure assuming a static envelope, or (given the ongoing evolution
of the nova on time-scales of maybe weeks or months) a sequence of
quasi-static envelopes.

\begin{figure}
\begin{center}
\plotone{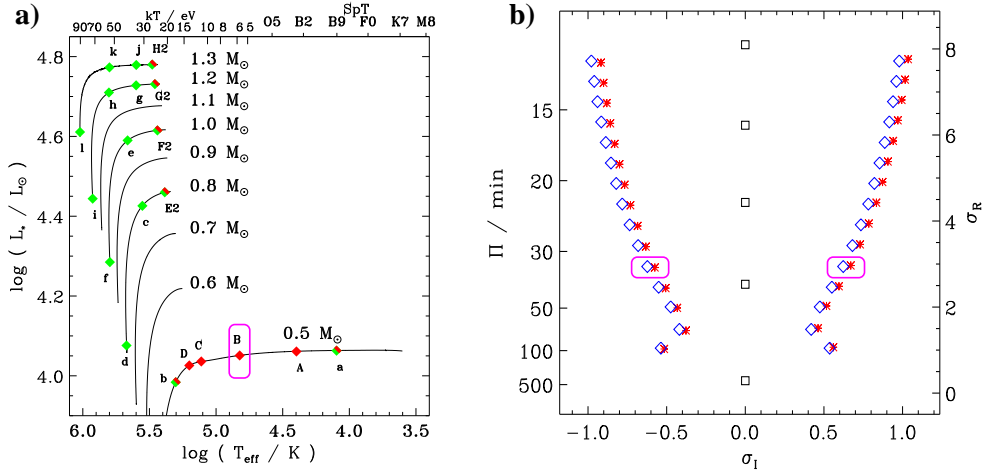}
\caption{a) The quasi-evolutionary sequences and the specific 20
models analysed --- 
b) Complex Eigenspectrum $(\sigma_{\rm R},\sigma_{\rm I})$ of model B
in different numerical treatment: squares adiabatic, diamonds NAR, 
asterisks non-adiabatic.}
\end{center}
\end{figure}

%%%%%%%%%%%%%%%%%%%%%%%%%%%%%%%%%%%%%%%%%%%%%%%%%%%%%%%%%%%%%%%%%%%%%%%%

\section{Models \& Results}

The background models for the linear analysis have been constructed by
a linear series of core plus envelope models, obtained via a shooting
method. Figure 1a (left panel) shows their position in the HR diagram. 
Along these curves a radial, non-adiabatic linear stability analysis
has been performed utilising the Riccati method (Gautschy \& Glatzel
1990). For the 20 models specifically marked I have also studied 
the non-linear evolution with adaptive-grid radiation hydro-code (RHD,
Dorfi \& Feuchtinger 1991).

One major result is plotted in Figure 1b (right panel), showing
the complex Eigenspectrum of model B
($M = 0.5 \, M_{\sun}, \log ( T_{\rm eff} / {\rm K} ) = 4.8$).
Frequencies $\sigma$ are plotted normalised with the free-fall 
time-scale in full non-adiabatic treatment and two different
approximations. 
Most striking are the enormous growth rates found, of the same order
as the free-fall time-scale. 
This is the case not only in model B, but for an extended range of
$T_{\rm eff}$. 
In contrast to classical pulsators on the instability strip (e.g.\ RR
Lyr stars) here the adiabatic approximation cannot reproduce the
frequencies at all. However, a highly suitable approximation for such
a situation is the Non-Adiabatic Reversible approximation (NAR), also
used by Glatzel (1994) and Saio, Baker, \& Gautschy (1998) in their 
investigations into the nature of strange modes. One typical
characteristic is the occurrence of complex conjugate pairs of
Eigenfrequencies as apparent in Figure 1b. 

\begin{figure}
\begin{center}
\plotfiddle{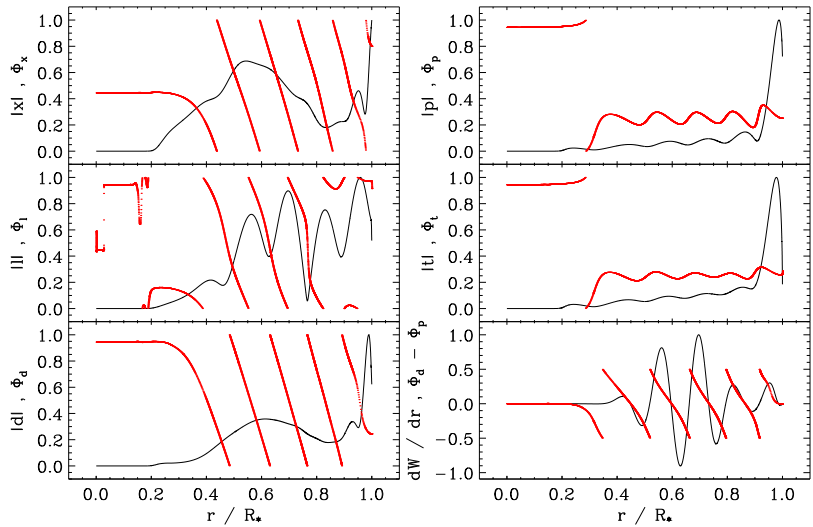}{5.7cm}{90}{45}{45}{160}{-10}
\caption{Set of fully non-adiabatic, complex Eigenfunctions for the
unstable mode of model B with 
$(\sigma_{\rm R},\sigma_{\rm I})=(2.92,-0.57)$, i.e.\ from the 5th
pair in Fig.~1b. In the first 5 panels, the relative Lagrangian
perturbations of radial displacement $x$, luminosity $l$, density $d$,
pressure $p$, and temperature $t$ are shown split into amplitude and
phase angle. The last panel shows the phase difference between density
and pressure perturbation together with the differential work integral.}
\end{center}
\end{figure}

NAR represents the Eigenmodes almost perfectly around the centre of
the instability region (model B), not only in frequency (Fig.~1b) but
also in terms of the Eigenfunctions.
These Eigenfunctions (cf.\ Fig.~2) represent complex standing, but in
their projection {\rm running} waves (in $x$, $l$, and $d$): outwards
for the unstable, and inwards for the stable mode of each pair.
In Figure 2, fully non-adiabatic Eigenfunctions are shown by
amplitude and phase, so perfectly complex conjugate pairs of waves
running inwards \& outwards would have identical arguments but 
negative phase in each of the Eigenfunctions. 
This is almost fulfilled and a mathematical requirement for NAR
solutions, with the exception of the luminosity perturbation $l$ which
is effectively set to zero in this approximation.

%%%%%%%%%%%%%%%%%%%%%%%%%%%%%%%%%%%%%%%%%%%%%%%%%%%%%%%%%%%%%%%%%%%%%%%%

\section{Interpretation \& Observational Consequences}

An improved local analysis with gradients on such models reveals
their nature as radiation-modified acoustic modes. It also allows to 
understand the spectrum found (e.g.\ in Fig.~1) to some degree.

The complementary analysis using the RHD code provides further 
characteristics of the modes and consequences in the non-linear regime. 
During the linear growth, and more so lateron, an additional energy
transport is observed, as packets of radiation energy are spiralling
outwards in density
perturbations (similar to a "screw conveyor").
This is likely to be responsible for the new "equilibrium" structure
of background model, which actually forms a curve on the HR diagram
which can significantly deviate from the initial position.

As a possibly related phenomena with non-radial behaviour of envelopes
under strong radiation fields, Shaviv (2001) has studied the
application of Super-Eddington winds on novae. In his model spatial
variations of opacity allow a larger radiation flux to pass, whereas
in the 1-dimensional case here only temporal modulations provide a
similar effect.

Although the project initially attempted to explain oscillation during
the so-called "transition phase" (Bianchini, Friedjung, \& Brinkmann
1992), a more probably scenario for direct observational evidence lies
in the predicted X-ray oscillations shortly before the turn-off of many 
novae (including the faster ones on more massive WDs, cf.\ Fig.~1a).

%%%%%%%%%%%%%%%%%%%%%%%%%%%%%%%%%%%%%%%%%%%%%%%%%%%%%%%%%%%%%%%%%%%%%%%%

\section{Conclusions}

It has become evident that static equilibrium models of novae after
visual maximum are subject to violent pulsational instabilities.
This finding (i) may suggests a possible link between mass loss in
novae and pulsations, but (ii) at least excludes a purely quasi-static 
evolution even for the lowest mass novae (possibly systems like V723
Cas and HR Del), where common mass loss mechanisms tend to fail.
For more massive WDs, only the very late phases of the outburst could be 
modelled. While they do also show instabilities, they still might
pulsate even during the non-static earlier part.

Classical novae form an excellent laboratory for pulsations under
radiation dominated conditions, so both fields will benefit from their
further detailed study. 

%%%%%%%%%%%%%%%%%%%%%%%%%%%%%%%%%%%%%%%%%%%%%%%%%%%%%%%%%%%%%%%%%%%%%%%%

%\medskip

\acknowledgments

This work was financially supported by the Swiss National Science
Foundation. 

%%%%%%%%%%%%%%%%%%%%%%%%%%%%%%%%%%%%%%%%%%%%%%%%%%%%%%%%%%%%%%%%%%%%%%%%

%%%%%%%%%%%%%%%%%%%%%%%%%%%%%%%%%%%%%%%%%%%%%%%%%%%%%%%%%%%%%%%%%%%%%%%%

\section*{Discussion}

{\it A. Cox~:} Is there any convection included in your calculations?\\[0.2cm]
{\it K. Schenker~:} I do not believe that convective effects are very
important in this case. These envelopes are strongly dominated by
radiation, and the radiative flux in the static models (which do
include convection via MLT) is indeed much higher that its convective
counterpart. So the linear analysis neglects the contribution of the
convective flux altogether, which is consistent with the RHD models
that do not contain convection at all.
\\[0.4cm]
{\it A. Cox~:} What is the temperature of the pulsating driving region?\\[0.2cm]
{\it K. Schenker~:} The driving temperature lies in the range of the
`iron peak' present in the new OPAL opacities, around 
$10^{5.2} \, {\rm K}$.


\begin{references}
\reference Bianchini, A., Friedjung, M., \& Brinkmann, W.~1992, \aap, 257, 599
\reference Dorfi, E.~A., \& Feuchtinger, M.~U.~1991, \aap, 249, 417    
\reference Gautschy, A., \& Glatzel, W.~1990, \mnras, 245, 154    
\reference Glatzel, W.~1994, \mnras, 271, 66
\reference Saio, H., Baker, N.~H., \& Gautschy, A.~1998, \mnras, 294, 622
%\reference Schenker, K.~1998, in {\it A Half Century of Stellar Pulsation
%Interpretations: A Tribute to Arthur N.~Cox}, ASP Conference Series 135, 116
\reference Schenker, K.~1999, PhD Thesis, Universit{\"a}t Basel
\reference Shaviv, N.~J.~2001, \mnras, 326, 126
\end{references}
\end{document}